\documentclass[notitlepage,aps,pra,reprint,twocolumn,longbibliography]{revtex4-2}
\usepackage{graphicx}
\usepackage{amsmath}
\usepackage{amssymb}
\usepackage{comment}
\usepackage[colorlinks, allcolors=blue]{hyperref}
\usepackage[all]{hypcap}
\usepackage[mathlines]{lineno}
\usepackage{braket}
\usepackage{wrapfig}
\usepackage{lipsum}
\usepackage{ulem}

\newcommand{\pref}[2]{\hyperref[#1]{\ref{#1}(#2)}}
\newcommand{\preff}[2]{\hyperref[#1]{\ref{#1}#2}}
\newcommand{\eqpref}[1]{\hyperref[#1]{(\ref{#1})}}

\newcommand{\squig}{{\raise.17ex\hbox{$\scriptstyle\sim$}}}

\begin{document}
	\title{Gray molasses cooling of $^{39}$K atoms in optical tweezers}
	\author{Jackson Ang'ong'a}
	\author{Chenxi Huang}
	\author{Jacob P. Covey}
	\email{jcovey@illinois.edu}
	\author{Bryce Gadway}
	\email{bgadway@illinois.edu}
	\affiliation{Department of Physics, University of Illinois at Urbana-Champaign, Urbana, IL 61801-3080, USA}
	\date{\today}

\begin{abstract}
Robust cooling and nondestructive imaging are prerequisites for many emerging applications of neutral atoms trapped in optical tweezers, such as their use in quantum information science and analog quantum simulation.
The tasks of cooling and imaging can be challenged, however, by the presence of large trap-induced shifts of their respective optical transitions.
Here, we explore a system of $^{39}$K atoms trapped in a near-detuned ($780$~nm) optical tweezer, which leads to relatively minor differential (ground vs. excited state) Stark shifts.
We demonstrate that simple and robust loading, cooling, and imaging can be achieved through a combined addressing of the D$_\textrm{1}$
and D$_\textrm{2}$
transitions.
While imaging on the D$_\textrm{2}$ transition, we can simultaneously apply $\Lambda$-enhanced gray molasses (GM) on the D$_\textrm{1}$ transition, preserving low backgrounds for single-atom imaging through spectral filtering.
Using D$_\textrm{1}$ cooling during and after trap loading, we demonstrate enhanced
loading efficiencies as well as cooling to low temperatures.
These results suggest a simple and robust path for loading and cooling large arrays of potassium atoms in optical tweezers through the use of resource-efficient near-detuned optical tweezers and GM cooling.

\end{abstract}
\maketitle
Neutral atom arrays have become a prominent platform for quantum information science. Most applications require repeated detection of atoms while keeping them cold. To reach temperatures near the motional ground state for alkali atoms in optical tweezers, designated sideband cooling along all three trap axes is required~\cite{Yu18,Thompson13,lorenz2020raman,govind19}.
Implementing such cooling in large arrays while preserving low-background detection via fluorescence from the same atomic transition remains challenging.

Recent work with alkaline earth atoms in optical tweezers~\cite{cooper18,norcia18,covey18} and atoms in an optical lattice~\cite{Cheuk2015PRL,Yamamoto_2016,Haller2015Sep} has demonstrated the advantages of using separate atomic transitions for the cooling and detection protocols, respectively, to enable versatile cooling near the motional ground state while simultaneously detecting atoms with low background~\cite{Haller2015Sep}. Such techniques have recently been extended to $^{39}$K atoms in optical tweezers~\cite{lorenz2020raman}, but they were complicated by large anti-trapping effects of the excited states at the trapping wavelength of 1064~nm, such that the trapping and light scattering protocols must be performed stroboscopically~\cite{lorenz2020raman,Hutzler_2017}.

Here, we implement a simple and robust cooling, trapping, and imaging scheme for $^{39}$K atoms using both their D$_\textrm{1}$ (770~nm) and D$_\textrm{2}$ (767~nm) transitions in optical tweezers of wavelength 780~nm. Minimal polarizability mismatch between the ground and excited states due to trap light at 780~nm enables efficient loading and robust in-trap cooling and nondestructive imaging (survival probability of $>$99$\%$) of atoms without much experimental overhead.
This configuration also gives a large polarizability for the ground state, enabling scalability to large arrays.
Moderate differential light-shifts due to the near-detuned optical tweezers allow us to
perform robust cooling by $\Lambda$-enhanced D$_\textrm{1}$ gray molasses (GM), which has been shown to be well-suited to scalability~\cite{Brow-GM,aliyu2021increasing}.
This approach also allows us to
demonstrate enhanced loading efficiencies ($\gtrsim$75$\%$) facilitated by
blue-detuned light-assisted collisions on the
D$_\textrm{1}$ transition~\cite{Grunzweig2010}.
These conditions suggest a robust route to loading large arrays of $^{39}$K atoms in optical tweezers.

\begin{figure*}[t!]
	\includegraphics[width=\textwidth]{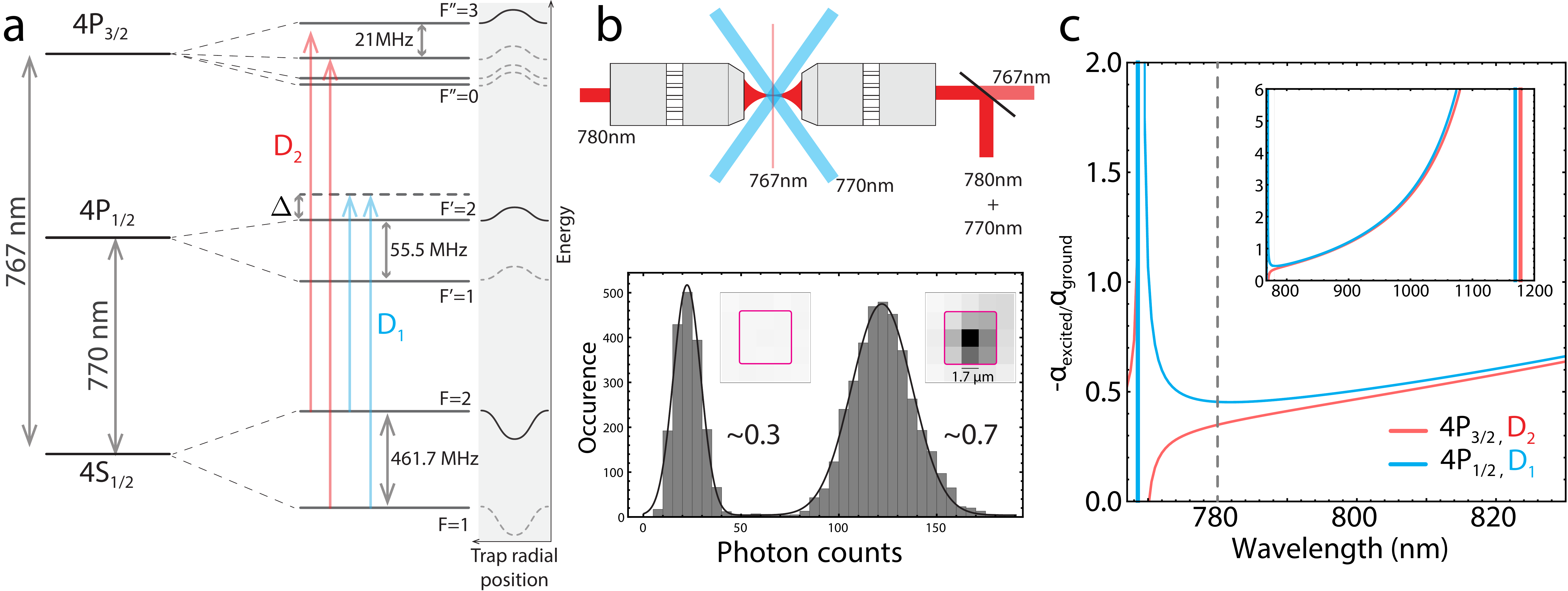}
	\caption{\label{FIG:fig1}
		\textbf{In-trap imaging of $^{39}$K in a 780~nm optical tweezer.}
		\textbf{(a) Left:}~$^{39}$K level diagram (not to scale). Gray molasses cooling using the D$_\textrm{1}$ $F=2 \rightarrow F'=2$ transition (blue) enables robust in-trap cooling while simultaneously performing fluorescence imaging on the D$_\textrm{2}$ transition (red).~\textbf{(a) Right:}~Illustration of trap-induced light shift for trap wavelength of 780~nm. Solid lines indicate the states relevant for imaging and cooling in our system.
		\textbf{(b)~Top:}~Sketch of the experimental setup. One high-NA objective delivers optical tweezer light (780~nm, dark red).  A small D$_\textrm{2}$ scattering beam (767~nm, light red) aligned perpendicular to the imaging axis is used to scatter photons that are collected using a second objective. D$_\textrm{1}$ gray molasses beams (770~nm, blue) are used to cool the atoms while performing D$_\textrm{2}$ imaging. Scattered fluorescence at 767~nm is collected through a second objective lens and imaged onto a camera, while the trapping and cooling light at 780~nm and 770~nm are removed through the use of narrow spectral filters.
		\textbf{(b)~Bottom:}~A histogram of photons collected from atoms loaded to an optical tweezer, collected over 5555 experimental shots under typical imaging conditions. The histogram demonstrates good detection efficiency ($>$99$\%$)
		and 68(1)$\%$ loading efficiency. The insets show the averaged image of shots determined to have zero atoms (left panel) and a single atom (right panel), with the pink square showing the region used for photon counting.
		This histogram is based on atoms loaded into an optical tweezer having a trap depth of $U/k_B = 1.1(1)$~mK and imaged for a duration of 50~ms.
		\textbf{(c)}~Influence of tweezer light shifts on the D$_\textrm{1}$ and D$_\textrm{2}$ transitions. We plot the excited state polarizabilities for the D$_\textrm{1}$ ($4P_{1/2}$, blue) and D$_\textrm{2}$ ($4P_{3/2}$, red) transitions, normalized to the ground state ($4S_{1/2}$) polarizability.
		The main panel shows the excited-to-ground state polarizability ratio for tweezer wavelengths near-detuned from resonance, with the experimental value of 780~nm noted by the vertical black line. Inset: The same quantities, but shown over a large range of tweezer wavelengths.
		}
\end{figure*}

\section{Experimental setup}
\label{exptsetup}

Starting from a bulk sample of laser-cooled $^{39}$K atoms~\cite{SuppMats}, we probabilistically load atoms into a single optical tweezer having 780~nm wavelength.
The trap laser light
is
passed through an acousto-optic modulator to enable power-stabilization and fast switching. The diffracted path is delivered to the atoms by a polarization-maintaining optical fiber. This light is filtered, shaped, and focused down to approximately 1.0(1)~$\mu$m by a high-NA objective lens (Mitutoyo G Plan Apo 50X; NA~$=$~0.5).
We operate with a typical trap beam power of 1.1~mW at the
focus,
relating to a typical trap depth of 1.1(1)~mK, as calibrated by the differential (excited vs. ground state) light-shift to the D$_\textrm{1}$ transition.
To note, this calibration assumes that the atom resides at the trap minimum (\textit{i.e.}, an assumption of zero temperature), and we expect the systematic calibration error due to residual motion is below $10\%$.
From our trap depth measurements and assuming a circular symmetry of the beam profile, we infer a trap waist of 1.0(1)~$\mu$m. This value is consistent with direct measurements of the beam at the focal plane.

Due to the compressed level structure of the $4P_{3/2}$ excited states for $^{39}$K [shown in Fig.~\ref{FIG:fig1}(a)], sub-Doppler temperatures achievable using bright molasses cooling on the D$_\textrm{2}$ transition are typically high compared to other alkali species. As a result, cooling and imaging atoms in traps having mK-level depths exclusively using D$_\textrm{2}$ bright molasses can be challenging.
The requisite cooling can alternatively be provided by $\Lambda$-enhanced gray molasses (GM) on the D$_\textrm{1}$ transition~\cite{Salomon_2013}, which has limiting temperatures far below those of the $^{39}$K D$_\textrm{2}$ bright molasses.
The combination of imaging on the D$_\textrm{2}$ transition (767~nm) and cooling by GM on the D$_\textrm{1}$ transition (770~nm) in principle provides a simple means to achieve robust cooling and imaging for $^{39}$K and other light alkalis, with the added benefit that cooling light can be spectrally filtered from the collected fluorescence.

To generate fluorescent scattering at 767~nm, we utilize a single small ($\sim$1~mm diameter) scattering beam that contains components near both the D$_\textrm{2}$ cycling and repump transitions. Its small size and alignment along a path that is perpendicular to the imaging axis, as shown in Fig.~\ref{FIG:fig1}(b), helps to minimize the amount of 767~nm stray scattering that enters the imaging system.
The 767~nm fluorescence light is collected by an objective that is identical to the one used for tweezer preparation, also shown in Fig.~\ref{FIG:fig1}(b).
This objective is part of a multi-lens imaging system that yields a net $\times$9.4 magnification, as imaged onto our EMCCD camera (Andor iXon Life).

While imaging via D$_\textrm{2}$ fluorescence, we simultaneously cool the atoms via $\Lambda$-enhanced GM on the D$_\textrm{1}$ transition. This combination enables relatively simple and robust trapping, cooling, and imaging of individual $^{39}$K atoms, as demonstrated by the histogram of collected fluorescence shown in Fig.~\ref{FIG:fig1}(b), which shows good detection fidelity ($>$99$\%$~\footnote{We fit both the zero- and one-atom peaks of the photon count histogram to Gaussian functions.
We then integrate the weight of the normalized zero-atom distribution that falls above a set threshold value of 55 counts.
Similarly, we integrate the weight of the normalized one-atom distribution falling below this threshold.
Each of these quantities are well below 1$\%$}) and an enhanced loading probability (68(1)$\%$) for a modest 50~ms imaging duration.

Here, the simplicity and robustness of our setup benefits from the use of optical tweezer light with a relatively small 10~nm detuning from resonance.
For more general conditions, the state-dependent light shifts (ac Stark shifts) generated by the optical tweezer can challenge schemes for cooling and imaging. While $^{39}$K does not permit any ``magic trapping'' (identical polarizabilities for ground and excited states) conditions for near infrared tweezer light, the effects of the state-dependent light shifts are less severe when operating at wavelengths not too far from resonance.
Figure~\ref{FIG:fig1}(c) presents calculations of the ratio of excited-to-ground state polarizabilities for the D$_\textrm{1}$ (blue) and D$_\textrm{2}$ (red) transitions due to tweezer light of varying wavelength.
Roughly speaking, the differential light shifts
(for a modest trap of mK-level depth)
are not too severe for trapping wavelengths ranging from roughly 774~nm to $\sim$900~nm. Outside of this range, one would expect that large inhomogeneous differential light shifts and the strongly anti-trapping nature of the excited states could present a serious challenge to the loading, cooling, and imaging of atoms.
Such issues can be overcome through staggered chopping of the cooling and trapping light~\cite{EdgeChop,Hutzler_2017}, as has recently been demonstrated for $^{39}$K atoms in 1064~nm wavelength optical tweezers~\cite{lorenz2020raman}.
However, continuous cooling by D$_\textrm{1}$ GM presents a simple and robust alternative when operating at near-infrared tweezer wavelengths below $\sim$900~nm.

To enable imaging with single-atom number resolution, it is highly beneficial to reduce the amount of background light that reaches the camera. Because our cooling and scattering beams have distinct wavelengths, we can use commercial color filters to remove the D$_\textrm{1}$ cooling light
(as well as the 780~nm trapping light)
from the imaging path.
This approach is relatively straightforward to implement, with no spatial filtering outside of the use of lens tubes along the imaging path.
However, particular to our present implementation,
which utilizes angled filters,
a small amount of residual background light enters our imaging system where beam tubes are not present.
This motivates us to work at rather large scattering rates ($\sim$140~kHz) to acquire signal faster than the background.

We note that this technical issue could be improved by better spatial filtering or by imaging on the far-separated 405~nm ($4S \leftrightarrow 5P$) transition, and that such an issue would be more easily avoidable for species with larger fine-structure splittings~\cite{Brow-GM}.
Still, as demonstrated by the histogram of Fig.~\ref{FIG:fig1}(b) and by the results that follow, we are able to achieve efficient and robust loading, trapping, cooling, and imaging of individual $^{39}$K atoms.

\section{Results and Discussion}
\label{results}

\subsection{Nondestructive imaging}

For the imaging of single atoms in optical tweezers, one would like to achieve a high discrimination between cases of having zero or one atoms per trap. We have demonstrated this capability in Fig.~\ref{FIG:fig1}(b), for modest imaging durations of 50~ms and for a trap depth of 1.1(1)~mK. In addition, it is also desirable to achieve low atom loss during the imaging. This is generally important for applications in quantum information science and analog simulation. Moreover, specific to the preparation of large defect-free fiducial states by the sorting of atoms to distill vacancies from tweezer arrays~\cite{Endres10,Barredoaah3778,Barredo17}, one would like to achieve a low loss probability during an image cycle. That is, the faithful sorting of atoms in an array based upon an initial image of trap occupations will only succeed if one retains all of the atoms observed in said image. Thus, to scale to arrays of hundreds of atoms, one should ensure survival probabilities in excess of $99\%$.

\begin{figure}[t!]
	\centering
	\includegraphics[width=0.97\linewidth]{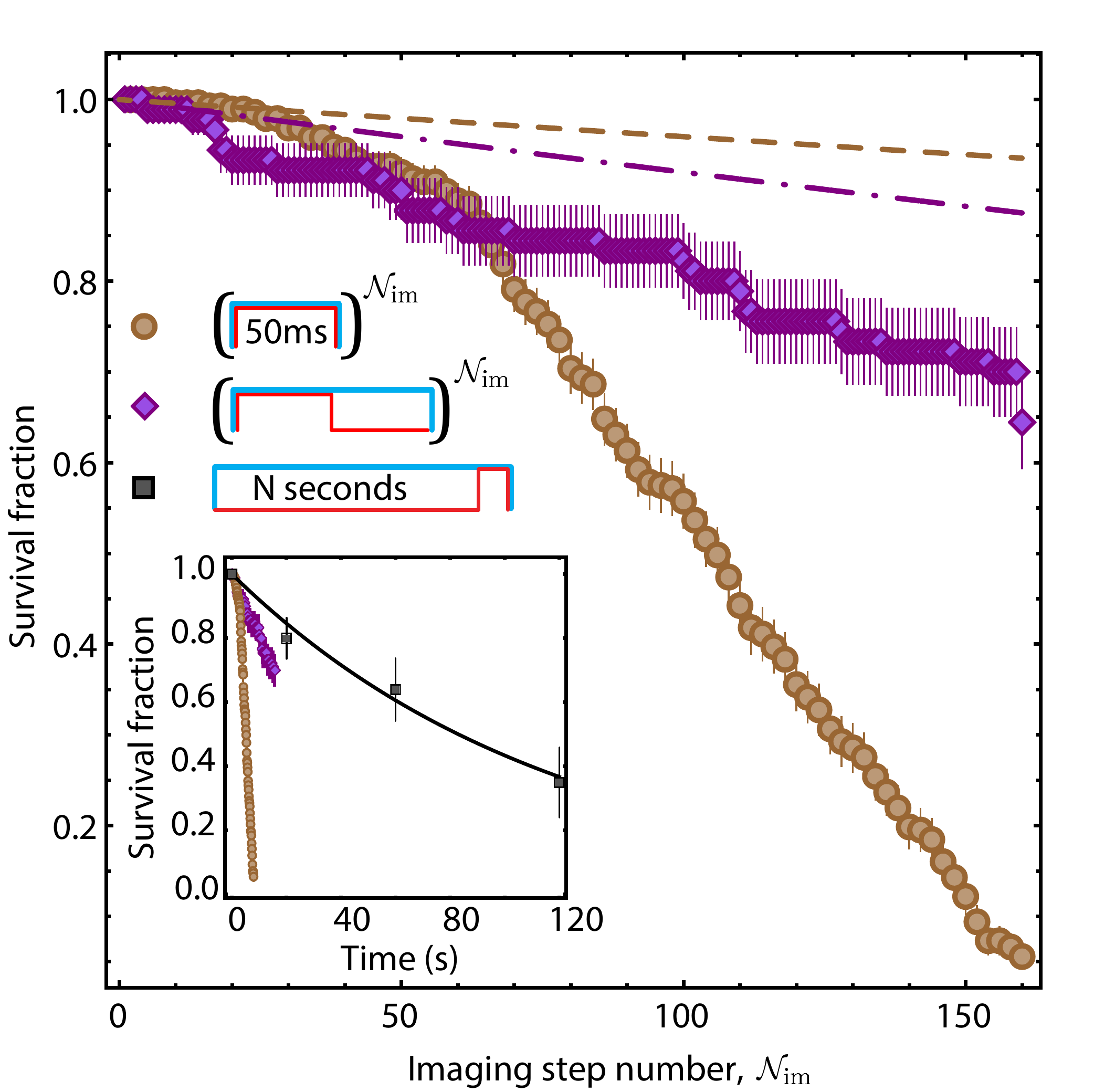}	
    \caption{\label{FIG:fig2}
	\textbf{Nondestructive imaging.}	
	Plot of the survival fraction of post-selected atoms in a 1.1(1)~mK-deep trap under repeated imaging (brown circles) or repeated imaging with interleaved fixed GM cooling (purple diamonds), shown as a function of the number of imaging blocks, which are defined in the legend. For each case, the image for each imaging block is acquired during a 50~ms-long period.
	Inset: Survival fraction as a function of time for the two imaging modalities shown in the main panel, as well as for the condition of simply holding in the trap with fixed D$_\textrm{1}$ GM and no D$_\textrm{2}$ scattering (black squares). The fit decay curve for the black data is reproduced in the main panel, with a conversion to the number of imaging blocks for the two modalities, shown as the brown dashed and purple dash-dotted lines.
	All error bars represent the standard error of the mean.
	}
\end{figure}

We begin by investigating the survival dynamics of atoms under the imaging conditions used to produce the histogram displayed in Fig.~\ref{FIG:fig1}(b).
Cooling by D$_\textrm{1}$ molasses is performed simultaneously while the small D$_\textrm{2}$ scattering beam is applied to the atom. These conditions are kept in place for a total of 8.05~s, with the first 50~ms relating to an initial imaging step
utilized for post-selection.
Then, conditioned on the presence of atom in this initial shot, we assess the probability for loaded atoms to survive after a varying number of imaging cycles $\mathcal{N}_\textrm{im}$. Specifically, while maintaining ``imaging conditions'' continuously for 8~s after the initial image, relating to a total of 160 imaging time steps, we record an image every other cycle, clearing the EMCCD array between successive shots.
The survival probability under such conditions (brown circles) is plotted in Fig.~\ref{FIG:fig2} as a function of the number of imaging cycles.

For a fixed loss rate $p_{\textrm{loss}}$ per imaging time step, one expects
an exponential decay of the survival probability, with a $1/e$ value reached for $\mathcal{N}_\textrm{im} \approx 1/ p_{\textrm{loss}}$.
Our data, however, displays a super-exponential decay over long times. This suggests that the loss rate is not constant, but increases as the atom is imaged, due to the finite trap depth and a heating rate that slightly exceeds the rate of cooling~\footnote{Owing to the differential trap light shifts and the blue-detuning of our D$_\textrm{2}$ scattering beam, this effect is further exacerbated by an increase in the scattering rate as the atoms are heated away from the trap minimum}.
Still, we can make some observations about the imaging survival probability based on this data.
First, our measured survival probability of $>$50$\%$ after 100 imaging time steps directly implies an initial survival probability of at least $99\%$ per image.
Second, an exponential fit to only the short-time data ($\mathcal{N}_\textrm{im} \leq 20$) suggests a survival probability of roughly $99.93\%$ ($p_{\textrm{loss}} \approx 0.0007$) per imaging step, promising for the production of defect-free arrays and for applications in quantum information science and analog simulation.

\begin{figure*}[t!]
	\includegraphics[width=\textwidth]{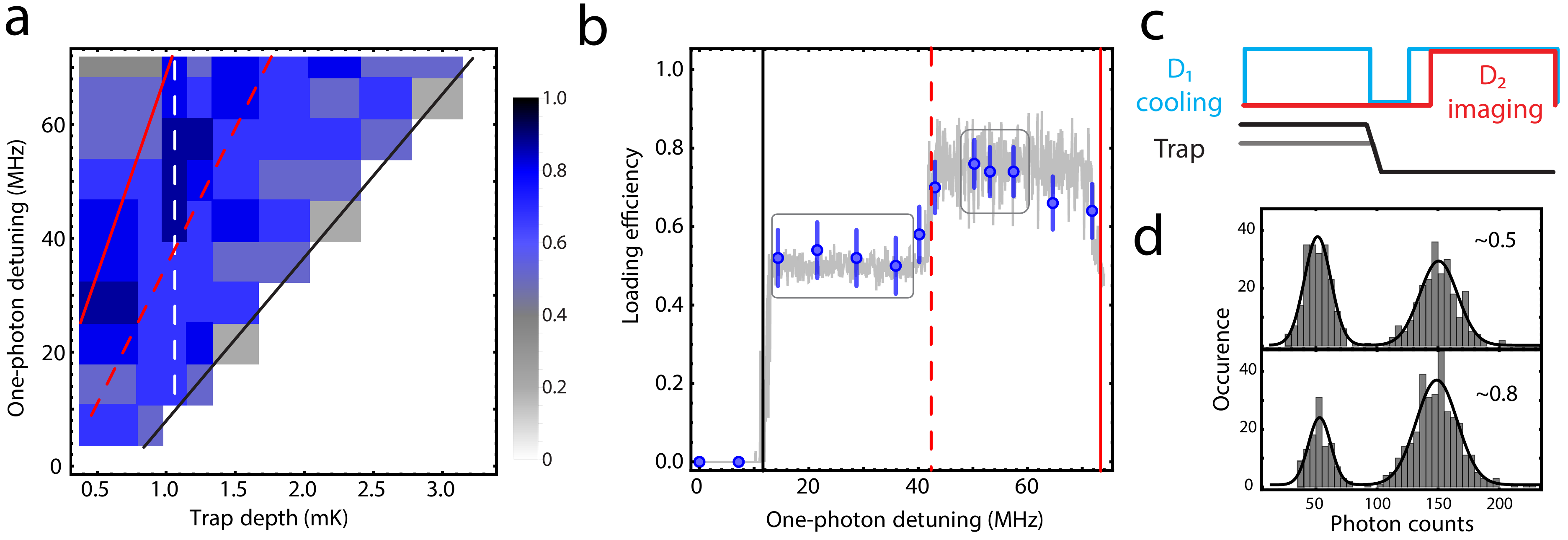}
	\caption{\label{FIG:fig3}
		\textbf{Gray molasses-aided loading of an optical tweezer.}
		\textbf{(a)}~Loading efficiency at different trap depths as a function of $\Lambda$-enhanced GM one-photon detuning. The black line indicates a differential linear light shift of the $F=2 \rightarrow F'=2$ transition induced by increasing trap depth. Above this line, cooling beams are blue-detuned and D$_\textrm{1}$ GM cools atoms in the trap. The
		red dashed (solid) line relates to an additional GM detuning by 1$\times$ (2$\times$) the trap depth ($U/h$).
		\textbf{(b)}~Loading efficiency taken at a fixed loading trap depth of 1.1(1)~mK [vertical white dashed line in (a)] as a function of the GM one-photon detuning $\Delta$. Below 12~MHz no loading takes place because cooling beams are red-detuned from the light-shifted D$_\textrm{1}$ transition. The loading efficiency jumps to roughly 50$\%$ once cooling beams are blue-detuned (black solid line). At one trap depth of additional detuning (red dashed line), the amount of energy absorbed by an atom pair during collision is only enough to eject one atom, thus increasing the loading efficiency.
		At two trap depths of additional detuning, the excess photon energy can eject both atoms from a colliding pair, presenting an upper edge for enhanced loading.
		The gray line corresponds to a Monte Carlo best-fit simulation incorporating atom pair loss during loading.
		\textbf{(c)}~Experimental sequence for GM loading an optical tweezer and subsequent atom imaging, as detailed in the text.
		\textbf{(d)}~Histograms of photon counts acquired under conditions of standard [$53(2)\%$, from lower four encircled points in (b)] and enhanced [76(2)$\%$, from upper three encircled points in (b)] atom loading.
	}
\end{figure*}

We now avoid the runaway loss of atoms and the associated super-exponential decay by simply including an interleaved cooling block between each imaging step~\cite{covey18}.
This added cooling step has the same duration as the imaging steps, and has the same fixed D$_\textrm{1}$ cooling conditions as during imaging. That is, we simply turn off the D$_\textrm{2}$ scattering beam during these cooling periods. The camera is exposed during the first $50$~ms, while scattering, and then cleared during each $50$~ms cooling block. The survival dynamics under this interleaved cooling protocol are also shown in Fig.~\ref{FIG:fig2} (purple diamonds), with the data taken over 16~s after the initial image.
Comparing to the case of continuous imaging, we primarily find that introducing interleaved cooling leads to a large enhancement of the survival probability after many imaging cycles, $\mathcal{N}_\textrm{im} \gtrsim 100$.~This is consistent with the expectation that the average cooling rate should exceed the average heating rate for the interleaved cooling protocol.
The survival data for interleaved cooling is also more globally consistent with a fixed loss probability per imaging cycle, with a value of $\sim$99.8$\%$ ($p_{\textrm{loss}} \approx 0.002$) suggested by an exponential fit to this full data set (purple diamonds).

Ultimately, aside from runaway processes and technical issues, the loss probability per imaging cycle should be set by the imaging cycle duration and some limiting loss rate set by background gas collisions, residual thermal loss under the steady-state (or cycle-averaged) cooling conditions, or a combination thereof.
We assess such limits for this system by measuring the survival fraction as a function of time under continuous D$_\textrm{1}$ cooling in the trap of depth $U/k_B = 1.1(1)$~mK (inset, black squares). An exponential fit to this data indicates a lifetime of $119.9\pm6.4$~seconds, limited by either vacuum or thermal loss under the steady-state D$_\textrm{1}$ conditions.
For comparison purposes, we plot the exponential decay from the fit to these conditions, converted to the number of imaging cycles that could be performed under both continuous imaging (brown dashed line) and interleaved-cooling imaging (purple dash-dotted line), in the main Fig.~\ref{FIG:fig2}. To note, this limiting loss rate appears to be roughly consistent with the short time survival probability for both the continuous imaging and interleaved cooling cases.

\subsection{Enhanced loading}

Single atom preparation in optical tweezers typically involves overlapping tweezers with a 3D~MOT and relying on inelastic light-assisted collisions induced by red-detuned molasses. Such collisions lead to pairwise loss from the trap, which projects the trap occupancy to 0 or 1, achieving parity projection and distilling out higher occupancies~\cite{Depue99PRL,Schlosser2002PRL, Schlosser2001Jun}.
In contrast to red-detuned light-assisted collisions, blue-detuned light can increase tweezer loading efficiency above 50$\%$ by preferentially expelling the higher-energy atom from a colliding pair~\cite{Grunzweig2010}.

We investigate loading of the 780~nm tweezer using $\Lambda$-enhanced GM, which involves blue-detuned D$_\textrm{1}$ laser light. Our procedure [Fig.~\ref{FIG:fig3}(c)] begins by loading a 3D~MOT and performing sub-Doppler cooling as described in Sec.~\ref{supp-expsystem}. The D$_\textrm{1}$ GM beams are kept on for 150~ms to load atoms into a variable-depth tweezer. We then turn off the GM beams for 40~ms to allow untrapped atoms to fall under gravity. During this time, the tweezer depth is linearly ramped from its initial ``loading'' value to a final fixed depth of 0.7(1)~mK.
After the 40~ms ``drop'' step, with only tweezer-trapped atoms remaining, we perform clean-up parity projection by turning the D$_\textrm{1}$ GM beams back on for 10~ms to induce pairwise loss. This step of parity projection by D$_\textrm{1}$ GM also serves to cool atoms prior to imaging.
We then image the atoms by turning on the D$_\textrm{2}$ scattering beam for 150~ms while simultaneously cooling by D$_\textrm{1}$ GM. To note, because of the lower trap depth in this case as compared to the conditions of Fig.~\ref{FIG:fig2}, we operate with a reduced scattering rate and a correspondingly longer imaging period.

We now explore the atom loading under different loading conditions, varying the initial trap depth $U$ as well as the one-photon detuning, $\Delta$, of the D$_\textrm{1}$ GM light. For each case, parity projection and imaging occur at the fixed final depth of 0.7(1)~mK.
For each set of loading conditions ($U$ and $\Delta$), we determine the single-atom loading probability by repeating the imaging procedure a minimum of 50~times.
As shown in Fig.~\ref{FIG:fig3}(a,b), we find that atoms are primarily loaded when $\Delta$, the one-photon detuning of the D$_\textrm{1}$ light from the free-space $F=2 \rightarrow F'=2$ transition, is positive (blue detunings).
This is consistent with the general dependence of D$_\textrm{1}$ GM cooling temperatures on the one-photon detuning, as previously reported in free space~\cite{Salomon_2013}.
We further observe that an additional blue-detuning of the GM light is required to load atoms in tweezer traps
of increasing depth~\cite{RegalGray}.
This compensates for the differential light shift $\Delta_{ge}$ produced by the 780~nm tweezer, which relates to 1.45~$U/h$, with $U$ the ground state trap depth ($h$, Planck's constant).

This differential light shift [$\Delta_{ge}$, as derived from the polarizability curves of Fig.~\ref{FIG:fig1}(c)] gives rise to the phenomenological black line shown in Fig.~\ref{FIG:fig3}(a), which denotes the onset of atom-loading.
To note, this measured shift of the D$_\textrm{1}$ GM loading edge with varying optical tweezer power and depth, which by construction has a slope of 29~MHz/mK, serves as our primary calibration of the ground state trap depth $U$.

A detailed view of the loading dependence on the D$_\textrm{1}$ detuning can be found by looking along a vertical cut (vertical white dashed line) of Fig.~\ref{FIG:fig3}(a).
The loaded atom fraction along this vertical cut,
taken at a loading trap depth of 1.1(1)~mK,
is plotted in Fig.~\ref{FIG:fig3}(b).
At zero detuning, no atoms are loaded, as the tweezer-induced light shifts cause the D$_\textrm{1}$ light to heat the atoms. For increasing detuning, we first observe a jump in the fraction of loaded atoms from 0 to roughly 50$\%$ at a moderate positive $\Delta$ value, with D$_\textrm{1}$ GM cooling enabled when the trap's differential light shift is overcome.
Here, the observed 50$\%$ loading is consistent with the application of parity projection prior to imaging.

\begin{figure}[t!]
	\includegraphics[width=0.95\columnwidth]{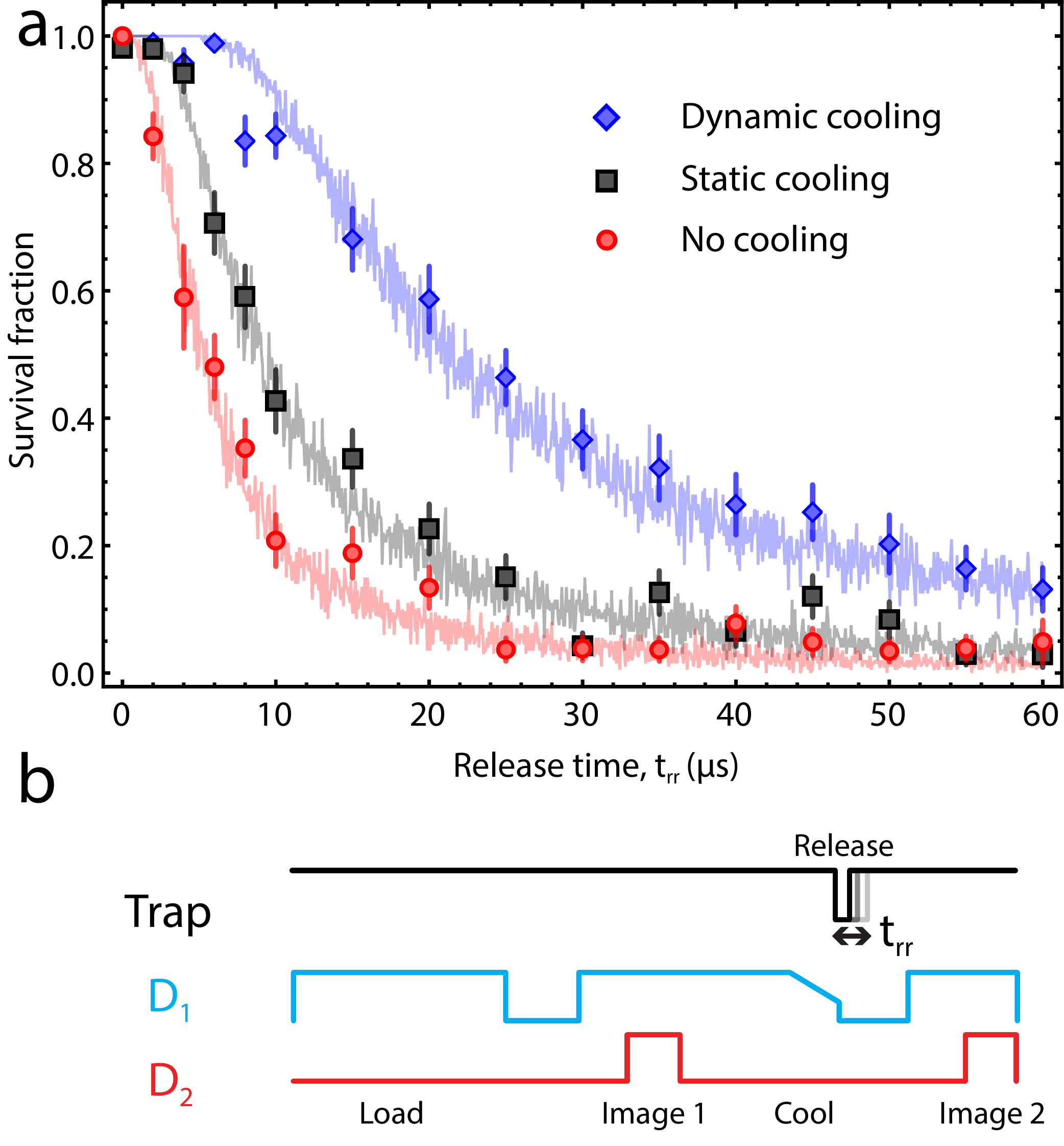}
	\caption{\label{FIG:fig4}
		\textbf{Temperature measurement by release and recapture.}
		\textbf{(a)}~Release-and-recapture survival probability of atoms released from an optical tweezer trap.
		We measure the fraction of pre-imaged atoms that are recaptured after releasing the atoms, by suddenly turning off the trap from a depth of $k_B \times 1.1(1)$~mK, and turning the trap back on after a variable delay time.
		We consider three different cooling scenarios prior to release: no additional cooling after pre-imaging~(red circles), a 100~ms block of cooling under fixed D$_\textrm{1}$ GM conditions~(black squares), and fixed cooling followed by a linear ramp of the D$_\textrm{1}$ GM to a lower final power~(blue diamonds).
		The points are data, while the shaded regions relate to Monte Carlo simulations used to infer the atomic temperature from the recapture fraction curves.
		Based on these comparisons, we extract temperatures of $\sim$130~$\mu$K, $\sim$60~$\mu$K, and $\sim$20~$\mu$K for the no cooling, static cooling, and dynamic cooling cases, respectively.
		The data error bars correspond to one standard error of the mean.
        \textbf{(b)}~Timing diagram for the trap light, D$_\textrm{1}$ cooling light, and D$_\textrm{2}$ scattering light, with the D$_\textrm{1}$ diagram relating to the dynamic cooling scenario.
		}
\end{figure}

This 50$\%$ loading is then seen over a range of increasing $\Delta$ values, until we observe another jump (red dashed line) to enhanced loading values in excess of 50$\%$.
We observe single-atom loading probabilities up to $76(2)\%$ at 1.1(1)~mK, as evaluated over 110 experimental shots.
The onset of enhanced loading is associated with the condition that, for pairs of atoms in the trap, one atom (having a greater lab-frame energy) may be preferentially ejected during a light-assisted collision by the excess photon energy deposited to the atom pair.
The onset of this condition relates to the red dashed line appearing in Fig.~\ref{FIG:fig3}(a),
which has a slope of $49\ \mathrm{MHz/mK}$ given by $\Delta_{ge} + U/h \sim 2.45 \ U/h$.
As $\Delta$ increases further, and along with it the excess energy imparted to atom pairs during light-assisted collisions, there is eventually enough energy to eject both atoms, pairwise, from the trap. This again results in parity-projection and a nominal loading rate of 50$\%$.
This upper threshold for enhanced loading is denoted by the solid red line in Fig.~\ref{FIG:fig3}(a,b), which has a slope of $69\ \mathrm{MHz/mK}$ given by $\Delta_{ge} + 2 \ U/h \sim 3.45 \ U/h$.

In practice, we find the enhanced loading conditions in the intermediate detuning range to be quite robust, routinely observing loading probabilities $\sim$70$\%$. Such enhanced trap loading probabilities can be useful, \textit{e.g.}, for reducing the resources and laser power overhead required to perform efficient sorting for the achievement of large defect-free tweezer arrays~\cite{RegalGray,aliyu2021increasing}.

\subsection{Gray molasses cooling}

Having demonstrated that the use of D$_\textrm{1}$ GM helps to enable low-loss, high-fidelity imaging as well as enhanced tweezer loading, we now investigate how well D$_\textrm{1}$ GM performs with respect to cooling tweezer-trapped $^{39}$K atoms. The reduction of residual atomic motion in optical tweezers is beneficial for a range of applications in quantum science. While Raman-sideband cooling is very well suited to achieving high trap ground state occupation~\cite{kaufman12,Thompson13} and has recently been demonstrated for $^{39}$K atoms~\cite{lorenz2020raman}, cooling by a simple and robust method like D$_\textrm{1}$ GM could still be useful for many purposes. For example, for quantum simulation studies based on Rydberg atoms released from optical tweezer arrays, where cooling prior to release helps to increase achievable free-space evolution times and reduce disorder in the inter-atom distances and interaction energies~\cite{Brow-Analysis}, the specific achievement of ground-state cooling is not as crucial as compared to, \textit{e.g.}, experiments involving coherent inter-tweezer tunneling~\cite{Jochim-tunnel,Kaufman-tunnel}.

Here, we assess cooling of atoms by performing measurements of the probability to recapture atoms following a sudden release from the trap and free expansion~\cite{Chu-molasses,Lett-molasses}.
Specifically, after loading atoms into a 1.1(1)~mK-deep trap and performing an initial image to verify occupancy, we suddenly extinguish the trap, wait a variable release time,~$t_{\mathrm{rr}}$, and then suddenly turn the trap back on and attempt a second image.
By comparing to Monte Carlo simulations, this release-and-recapture method allows for an estimate of the atom temperature, with particular sensitivity to the radial temperature~\cite{Tuchendler08}.

We now investigate the release-and-recapture survival dynamics under three different cooling scenarios.
(1)~No cooling: Following the initial image (50~ms in duration, same conditions as utilized in Fig.~\ref{FIG:fig2}), we simply release the atom with no additional cooling applied.
(2)~Static cooling: After taking the first image a static cooling block is included where D$_\textrm{1}$ GM is applied for 100~ms at a fixed power, under the same conditions as optimized for low-loss imaging. (3)~Dynamic cooling: Here we append an additional dynamical cooling block to the 100~ms-long static block, during which the D$_\textrm{1}$ GM laser power is ramped down to a lower value over 20~ms before the atom is released.
The dependence of release-and-recapture survival probability for these three cases are shown in Fig.~\ref{FIG:fig4} as a function of release time.

In the absence of cooling (red circles), we observe sharp decay of the survival probability upon release, with a survival rate of roughly 50$\%$ found after roughly 6~$\mu$s. A moderate increase in the achievable release times is observed upon the introduction of the static cooling block (black squares), with the survival probability at 6~$\mu$s increasing to $\sim$70$\%$.
For dynamic cooling, reminiscent of free-space gray molasses~\cite{GMo1,GMo2,Salomon_2013,Brow-GM}, we reduce the GM laser intensity to a third of its initial value (based on empirical optimization) over 20~ms to achieve an associated reduction of the atomic temperature.
As compared to the cases of no cooling or static cooling, we find that the survival dynamics under dynamic cooling (blue diamonds) dramatically improves, with essentially no loss found after 6~$\mu$s and a $1/e$ survival time of roughly 30~$\mu$s.

To estimate the temperature of the atom for the three cases we implement a Monte Carlo simulation~\cite{Tuchendler08,Alt03}. We consider a single atom in a 1.1(1)~mK trap with an initial position and velocity. The initial position is drawn from a Gaussian distribution with standard deviation given by $\sigma_z=\sqrt{k_B T / m \omega_z^2}$ and $\sigma_r=\sqrt{k_B T / m \omega_r^2}$
in the axial and radial directions, respectively. Here $\omega_i$ is the trap frequency in the $i^{\mathrm{th}}$ direction, $m$ is the mass of the atom, and $k_B$ is the Boltzmann constant. The initial velocity is also drawn from a Gaussian distribution with $\sigma_v=\sqrt{k_B T / m}$. The atom is evolved for some time $t$ using kinematic equations of motion and the final total energy is calculated. Specifically the total energy is a sum of the kinetic energy of the atom and the potential energy due to trap potential (which includes modification due to gravity). If the total energy of the atom is larger/smaller compared to the trap depth ($k_B \times1.1(1)$~mK), the atom is considered lost/re-captured and a trap occupation of 0/1 is assigned. We repeat the simulation 1000 times for each release time and average the outcomes to obtain a survival probability.

The temperature is now estimated by running the simulation for a varying range of $T$ to obtain survival probability curves as a function of release time, followed by a least-square analysis to find the best fit. Based on these fits we extract temperatures of roughly 130~$\mu$K, 60~$\mu$K, and 20~$\mu$K for the cases of no cooling, static cooling, and dynamic cooling, respectively.
For context, based on approximate estimated trap frequencies of $\omega_\perp / 2\pi = 180$~kHz and $\omega_\parallel / 2\pi = 30$~kHz (from the measured depth and inferred waist), the dynamic cooling temperature would relate to mean trap phonon occupations of roughly $n_\perp \sim 2$ and $n_\parallel \sim 11$ in the radial and axial directions, respectively.
While the lowest temperature we achieve is higher than what has been achieved through careful Raman-sideband cooling~\cite{lorenz2020raman}, the simplicity and robustness of implementation by D$_\textrm{1}$ GM could be of benefit for certain applications.

\section{Conclusions}
\label{conclusions}

We have demonstrated that D$_\textrm{1}$ gray molasses cooling can serve as a powerful resource for achieving enhanced loading, high-fidelity nondestructive imaging, and robust cooling of $^{39}$K atoms trapped in near-detuned (780~nm) optical tweezers. Of practical interest, the combination of D$_\textrm{1}$ GM with near-detuned tweezer light allows for a simple and robust implementation of trapping, cooling, and imaging. Moreover, the robustness of D$_\textrm{1}$ GM and the large ground state polarizability of the 780~nm tweezer light provide favorable conditions for the extension to larger arrays of tweezer-trapped atoms.
More broadly, bichromatic approaches to imaging and cooling atoms in optical tweezers should be widely applicable, and likely even easier to implement in many leading experiments based on heavier alkali species.

\section*{Acknowledgements}

We thank Nikolaus Lorenz and Christian Gross for helpful discussions.
We thank Cheeranjeev Purmessur, Ivan Velkovsky, Sai Paladugu, Garrett Williams, and Shraddha Agrawal for a critical reading of the manuscript.
We thank Garrett Williams, Michael Highman, Eric Meier, Tao Chen, and Zejun Liu for technical assistance and for valuable discussions.
This material is based upon work supported by the National Science Foundation under grant No.~1945031.

\bibliographystyle{apsrev4-1}
%

\clearpage

\section*{Supplementary Materials}
\renewcommand{\theequation}{S\arabic{equation}}
\renewcommand{\thefigure}{S\arabic{figure}}
\setcounter{equation}{0}
\setcounter{figure}{0}

\subsection{Experimental system}
\label{supp-expsystem}
Our experiments make use of a multi-chamber vacuum apparatus with two separated regions for magneto-optical trapping. We first, over several seconds, load a three-dimensional magneto-optical trap (3D~MOT) of $^{39}$K atoms in an intermediate trapping region. This intermediate 3D~MOT is continuously fed by a flux of atoms produced by a separate two-dimensional magneto-optical trap. After nearing full capacity, the 3D~MOT has its gradient extinguished and its atoms are transferred to a final 3D~MOT that resides in a glass science cell (rectangular quartz cell, with dimensions of 28~mm $\times$ 28~mm $\times$ 100~mm and a 4~mm wall thickness) roughly 0.5~m away. This transfer of the atoms is accomplished by a 200~$\mu$s-long pulse of near-resonant (blue-detuned from the D$_\textrm{2}$ transition) laser light that passes between the intermediate and final 3D~MOT chambers.
The laser light in the final 3D~MOT is turned on 20~ms after the atoms are pushed from the intermediate MOT to catch atoms and is allowed to Doppler-cool the atoms for 200~ms.

Our final MOT has several features particular to its use for single atom imaging and cooling in optical tweezers. It features laser light for addressing both the D$_\textrm{2}$ (767~nm) and D$_\textrm{1}$ (770~nm) transitions of $^{39}$K [see main text Fig.~\ref{FIG:fig1}(a)]. The
laser
light for the D$_\textrm{2}$ and D$_\textrm{1}$ transitions
is
combined by a fiber-based splitter-combiner.
To accommodate two high numerical aperture (NA) objectives for imaging and optical tweezer focusing, the final MOT features non-orthogonal beam paths, set apart by 120$^{\circ}$ in the horizontal plane [see main text Fig.~\ref{FIG:fig1}(b)]. A vertical MOT path provides cooling along the vertical direction.

After loading the D$_\textrm{2}$ MOT in the final chamber, we first compress the cloud by increasing the magnetic field gradient, and we then apply molasses cooling on the D$_\textrm{2}$ transition for 3.5~ms by turning off the gradient and changing the powers and detunings of the D$_\textrm{2}$ cycling and repumping light (near to the $F=2 \rightarrow F"=3$ and $F=1 \rightarrow F"=2$ transition lines, respectively).

We then further cool our bulk cloud of atoms by $\Lambda$-enhanced gray molasses (GM) on the D$_\textrm{1}$ transition~\cite{Salomon_2013}. Here, we apply D$_\textrm{1}$ cycling light (blue-detuned from the $F=2 \rightarrow F'=2$ transition) and repump light (blue-detuned from $F=1 \rightarrow F'=2$), which are set up in a $\Lambda$ configuration [see main text Fig.~\ref{FIG:fig1}(a)] with zero two-photon detuning.
Based on parameters optimized using Gaussian process implemented using an open-source software package (MLOOP~\cite{Wigley2016}), we routinely cool bulk gases of atoms down to $\approx$ 4~$\mu$K over 7.5~ms before loading the optical tweezer.

\begin{figure}[b!]
	\includegraphics[width=\columnwidth]{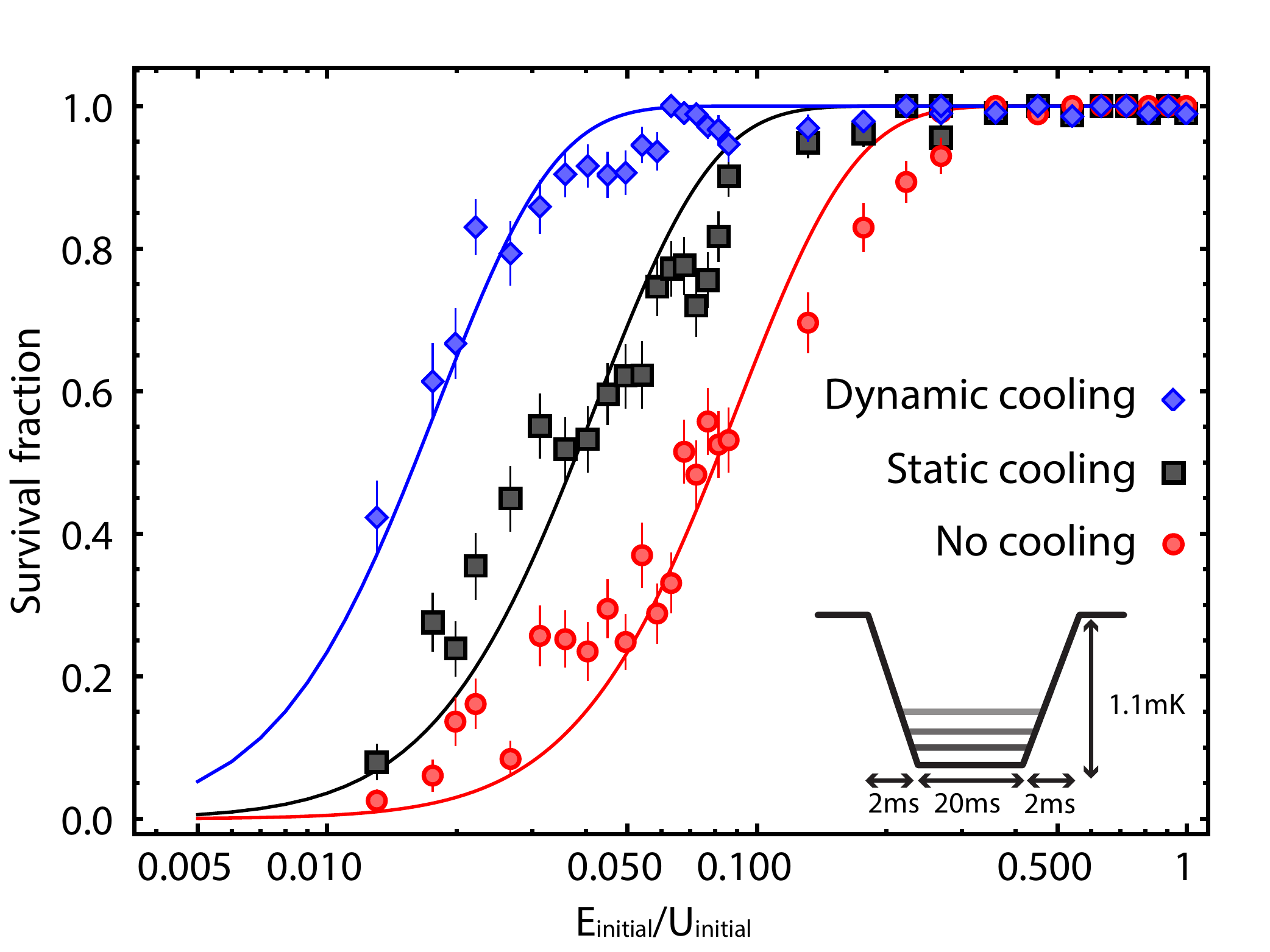}
	\caption{\label{figRampdown}
	\textbf{Temperature estimation by ramp-and-hold measurement.} We measure the fraction of atoms that remain in the trap after slowly turning down the trap power over 2~ms to variable depths, holding at that depth for 20~ms, and ramping back up over 2~ms to the initial depth ($U/k_B = 1.1(1)$~mK). The points are experimental data, and the lines relate to the best-fit (fit as a function of different trial initial temperatures) theory curves of survival fraction as based on integrating the action of the atoms throughout the ramp-down and comparing the resulting energy to the trap depth (following the procedure outlined in Refs.~\cite{Tuchendler08,Alt03}).
	The data error bars represent the standard error of the mean.
	The inset cartoon depicts the symmetric ramp-and-hold procedure. The red, black, and blue data/theory relate to the cases of no cooling, static cooling, and dynamic cooling, with the same preparation, imaging, and cooling procedures as performed for the data presented in Fig.~\ref{FIG:fig4} of the main text.
}
\end{figure}

\subsection{Temperature estimate by trap depth reduction}

As proposed in Ref.~\cite{Tuchendler08,Alt03}, the temperature of an atom can be measured through a rampdown technique where the trap depth is adiabatically lowered (by lowering trap power) and held for enough time for the atom to potentially escape before raising the depth again to image the atom.
In short, by adiabatically lowering to some minimum trap depth $U_\textrm{min}$, the atom escapes if its initial energy is sufficiently high such that the final energy after this ramp (considering integration of the action that describes the adiabatic rampdown procedure) is higher than $U_\textrm{min}$.
Based on this sensitivity to the initial energy, the measurement of survival after ramping down the trap to variable depths allows for the estimation of the atom temperature.

We experimentally implement this rampdown technique for the three different cooling conditions already investigated in the main text (through release and recapture). Specifically, following an initial imaging step, we implement one of the three cooling scenarios - no cooling, static cooling, or dynamic cooling. The trap depth is then lowered from 1.1(1)~mK to some minimum depth $U_{\mathrm{min}}$ over 2~ms, held at this depth for 20~ms, and then raised back up over 2~ms to a final depth of 1.1(1)~mK. A second image is then taken to determine whether the atom was retained or lost. To minimize any potential trap loss due to imperfect imaging, a 50~ms cooling block is added before each imaging time step. For each $U_{\mathrm{min}}$ explored, we take $\approx 180$~shots per data point. As shown in Fig.~\ref{figRampdown}, the measured survival fraction starts at 1 for $U_{\mathrm{min}}\sim 1$~mK and decreases for lower minimum trap depths.

In Fig.~\ref{figRampdown} we present the rampdown survival data, not directly as a function of the trap depth, but rather plotting as a function of a recalibrated quantity $E_\textrm{initial}/U_\textrm{initial}$~\cite{Tuchendler08,Alt03}. For decreasing trap depths, loss of atoms first occurs for the no cooling case (red circles), with 50$\%$ survival at around $U_{\mathrm{initial}} / 10$.
For the static cooling case (black squares),
we observe a clear influence of the added cooling,
as evidenced by the fact that survival reaches 50$\%$ at around $U_{\mathrm{initial}} / 20$ .
Finally, with the dynamic cooling step (blue diamonds), we find that the survival fraction remains relatively high for lower trap depths,
reaching 50$\%$ survival at roughly $U_{\mathrm{initial}} / 50$.

Qualitatively, these rampdown measurements are in good agreement with the release-and-recapture results of the main text. The release-and-recapture results revealed that static cooling lowered the temperature with respect to the no cooling scenario, and that dynamic cooling provided still further temperature reduction (\textit{i.e.}, $T_\textrm{dynamic} < T_\textrm{static} < T_\textrm{no cool}$).
A similar trend can be inferred from the rampdown data of Fig.~\ref{figRampdown}, based on the correspondence between the initial energy (temperature) and the achievable minimum trap depth prior to escape.

Following the procedure of Refs.~\cite{Tuchendler08,Alt03}, we now estimate the temperature of the atoms based on these rampdown measurements. Assuming a Maxwell--Boltzmann distribution for the initial atomic temperature, and based on a 1.1(1)~mK initial trap depth and an estimated trap waist of 1.0(1)~$\mu$m, fits to the survival data yield initial temperatures of $\sim$40~$\mu$K, $\sim$100~$\mu$K, and $\sim$170~$\mu$K for the dynamic, static, and no cooling scenarios. To note, these temperature estimates differ from those of the release-and-recapture method by almost a factor of two.

This quantitative disagreement could occur for various reasons. For one, the two methods are expected to have different relative sensitivities to the atoms' radial and axial temperatures, which may not coincide. That is, the atoms may have a low radial temperature, to which release-and-recapture is predominantly sensitive, but a higher axial temperature that would affect the rampdown survival.
Alternatively, the temperature estimates from these two procedures could be in apparent disagreement simply because of their unique dependencies on quantities like the trap depth and trap waist, combined with our uncertainties in these quantities.
As discussed in the main text, we have a statistical uncertainty in the trap depth from our calibration based on the differential light shift for the D$_\textrm{1}$ transition, which in turn also gives rise to an uncertainty in our inferred trap waist.
Combined, these 10$\%$-level uncertainties, along with potential deviations from our assumptions (such as circular symmetry of the trap beam), could also help to account for disagreement in the temperature estimates.

\end{document}